\documentclass[preprint,amsmath,amssymb,aps,pra,twocolumn,epsfig,showpacs,bibliography,lengthcheck]{revtex4-1}
\usepackage{graphicx}
\usepackage{amssymb, amsmath}
\usepackage{bm}
\usepackage{hyperref}
\usepackage{color}
\usepackage{mathtools}
\usepackage{float}
\begin{document}
\title{Spectral Asymmetry of Atoms in the van der Waals Potential\\ of an Optical Nanofiber}
\author{B. D. Patterson}
\author{P. Solano}
\author{P. S. Julienne}
\author{L. A. Orozco}
\author{S. L. Rolston}

%\email{lorozco@umd.edu}

\affiliation{Joint Quantum Institute, Department of Physics, University of Maryland and National Institute of Standards and Technology, College Park, MD 20742, USA}

\date{\today}

\begin{abstract}
We measure the modification of the transmission spectra of cold $^{87}$Rb atoms in the proximity of an optical nanofiber (ONF). Van der Waals interactions between the atoms an the ONF surface decrease the resonance frequency of atoms closer to the surface. An asymmetric spectra of the atoms holds information of their spatial distribution around the ONF. We use a far-detuned laser beam coupled to the ONF to thermally excite atoms at the ONF surface. We study the change of transmission spectrum of these atoms as a function of heating laser power. A semi-classical phenomenological model for the thermal excitation of atoms in the atom-surface van der Waals bound states is in good agreement with the measurements. This result suggests that van der Waals potentials could be used to trap and probe atoms at few nanometers from a dielectric surfaces, a key tool for hybrid photonic-atomic quantum systems.

 \end{abstract}
%\pacs{21.10.Gv,27.80.+w,32.10.Fn}

\maketitle

%=================================================================

\section{Introduction}
\label{sec:intro}

Optical nanofibers (ONF) are a new platform for the study and use of light-matter interactions \cite{Solano2017}. The small mode area of the evanescent field around the ONF waist leads to a large coupling between the ONF guided mode and atoms near its surface~\cite{LeKien2006}. This has triggered numerous studies where ONFs are used for sensing properties of other systems \cite {Brambilla2010,Zhang2010,Morrissey2013}, and in areas spanning atomic physics, quantum optics, and quantum information ~\cite{Lodahl2016,Solano2017}. 

All these studies and applications benefit from the large atom-light coupling available in ONFs, and larger coupling is achieved for atoms closer to the ONF surface. However, atoms that are near the ONF surface experience a van der Waals potential that attracts them towards the surface and shifts their energy levels. As such, characterization of the van der Waals potential produced by an ONF is essential for both practical applications and fundamental science \cite{Woods2016}. The atomic spectra can give information about state-sensitive potentials felt by the atoms, such as the van der Waals potential. Refs.~\cite{Sague2007,LeKien2007b,Russell2009,Minogin2010,Nayak2012a} are recent studies of spectra of atoms near the surface of an ONF and there are summaries in the reviews in Refs.~\cite{Solano2017,Morrissey2013}.

The pioneering work by Sagu{\'e} {\it et al.} \cite{Sague2007} measured the absorption spectra of Cs atoms near an ONF. They showed with numerical simulations that the presence of the nanofiber caused a significant decrease in the density of atoms near the surface because of the van der Walls attractive potential, while the scanning probe exerted attractive or repulsive forces on the atom from the large intensity gradient that depend on the sign of the detuning from resonance. They measured the evolution of the linewidth of the absorption as a function of probe power. Their lowest  width is about 6.2 MHz (higher than unperturbed-atom natural linewidth of 5.2 MHz) that increases to about 7 MHz at their highest probing power. Their complete model with the dipole forces and the surface interactions agrees well with all the measurement that focus on the linewidth rather than possible asymmetries.  

LeKien {\it et al.} and Russell {\it et al.} calculated spectra of atomic fluorescence near a nanofiber including the effects of surface-induced potentials, \cite{LeKien2007b,Russell2009} with Minogin {\it et al} \cite{Minogin2010} treating the manifestation of the van der Waals surface interaction in the spontaneous emission of atoms inside the evanescent mode of an optical nanofiber.

Nayak {\it et al.}~\cite{Nayak2012a} experimentally studied the fluorescence of atoms around an ONF by exciting them with a free space propagating beam and collecting the fluorescence that couples to the guided mode of the ONF. Their results show  significant asymmetries in the fluorescence spectrum and relate it to the desorption of atoms from the ONF. The shifts observed, in excess of 100~MHz, come from atoms in van der Waals bound states, for which they qualitatively show the predicted asymmetry. This observation raises the question of whether or not we can excite and probe atoms trapped in bound states of the van der Waals potential.

We look at the transmission spectrum of  atoms in the vicinity of a ONF by scanning the frequency of a weak resonant probe laser beam that propagates through the guided mode. We thermally excite atoms on the ONF surface in a controlled way by changing the nanofiber temperature with the power of an auxiliary non-resonant laser, which also propagates in the fundamental mode of the nanofiber. We observe asymmetric broadening of the natural line as a function of the heating laser power. A semi-classical phenomenological model of the transmission spectrum produced by atoms thermally populating van der Waals bound states qualitatively agrees with the observations. This result could open a way to perform quantum optics experiments using atoms trapped in the van der Waals potential, where the atom-light interaction is maximized for the system, and realize novel trapping geometries as proposed in Ref. \cite{Chang2014}

The paper is organized as follows.
Section~\ref{sec:system} outlines the nanofiber mode structure, the effect of the van der Waals potential in nearby atoms, and a model that describes the spectroscopic signal in our experiment. Section~\ref{sec:expt} provides a general overview of the apparatus.
Section~\ref{sec:res} presents the measured spectra and the extracted spectral asymmetries. The results are compared with a semi-classical model of thermal excitations of van der Waals bound states. This is followed by a discussion in Sec. \ref{sec:disc} and a conclusion in Sec. \ref{sec:concs}.

%=================================================================

\section{The system}
\label{sec:system}
The experiment relies on two main parts: a single mode ONF and a source of cold $^{87}$Rb atoms.
A magneto-optical trap (MOT), generated in the vacuum chamber from the residual atomic vapor pressure from a Rb dispenser, provides a source of atoms that couple to the evanescent field of the ONF guided mode.
The nanofiber can couple the light spontaneously emitted by the atoms into the guided mode and deliver it to a detector past the tapered zone of the ONF.

A standard optical fiber consists of a core of refractive index $n_\mathrm{core}$ and radius $a$, surrounded by a cladding with lower refractive index $n_\mathrm{clad}$ and radius $R$.  In our ONFs, $R$ is reduced to subwavelength dimensions by a flame-brush technique~\cite{Hoffman2014a}.  At these dimensions, the ONF can be considered as a single dielectric of index $n_\mathrm{ONF}$ = $n_\mathrm{clad}$, surrounded by vacuum, $n$ = 1, as the original fiber core becomes negligible.  The tapers connecting the standard fiber on the input and output side to the ONF waist have milliradian angles for adiabatic propagation~\cite{Ravets2013,Birks1992}.

%-------------------------------------------------------------------------------------------------------------------
\subsection{Mode structure and coupling}
\label{subsec:mode}
We use a single-mode nanofiber. The fundamental mode (HE$_{11}$) has an intensity profile outside of the ONF given by~\cite{LeKien2004,Snyder1983}
\begin{equation}
I(r) = \mathcal{E}^2\left[ K^2_0(qr) + u K^2_1(qr) + w K^2_2(qr)\right]\,,
\label{eq:intensity}
\end{equation}
where $\mathcal{E}^2$ is a proportionality constant; $K_i$ is the modified Bessel function of the second kind of order $i$; $u$ and $w$ are constants obtained from Maxwell's equations; $r$ is the distance from the center of the ONF; and $q = \sqrt{\beta^2-k^2}$ is the transverse component of the wavevector, where $\beta$ is the field propagation constant in the nanofiber, and $k = 2\pi/\lambda$ is the amplitude of the free-space wavevector.
The parameter $q$ describes the decay of the field in the radial direction.
The asymptotic expansion for large argument for the modified Bessel functions
$K_{l}(qr)\approx \sqrt{\pi/ 2 qr}e^{-qr}$
is a good approximation, for any order $l$ \cite{Olver2010}. Considering this, the radial dependence of the intensity of the evanescent electric field is
\begin{equation}
\label{eq:intensity-ap}
I(r)\approx \mathcal{E}^2 \frac{\pi}{2qr}e^{-2qr},
\end{equation}
This approximation shows that the evanescent field decays at a shorter distance than an exponential decay \cite{Solano2017}.

%-------------------------------------------------------------------------------------------------------------------

We define the emission enhancement parameter as the rate of spontaneous emission $\Gamma_{\text{1D}}$ that couples into the ONF mode divided by the natural atomic decay rate $\Gamma_{0}$ ~\cite{LeKien2006,Quan2009,Masalov2013,Solano2017},
\begin{equation}
\alpha \left ( r \right ) = \Gamma_{1\mathrm{D} }\left( r \right)/\Gamma_{ \mathrm{0} }\,.
\label{eq:coupling}
\end{equation}
$\alpha \left ( r \right )$ follows the spatial variation of Eq. (\ref{eq:intensity}), as stated by Fermi's golden rule. The emission enhancement is a unit-less parameter that characterizes the coupling strength between an atom and the ONF mode. Any signal coming from an atom emitting into the ONF mode will be scaled by this parameter.

%-------------------------------------------------------------------------------------------------------------------

\subsection{Surface potentials}
\label{subsec:surface}

The glass surface of the ONF interacts with an atom mediated by virtual photon exchange between induced dipoles. This gives rise to a van der Waals potential
\begin{equation}
\label{eq:vdWCP}
U (r) = - \frac{C_3}{r^3}\,,
\end{equation}
where $r$ is the radial distance from the ONF surface to the atom. $C_3$ is the van der Waals coefficient. For $^{87}$Rb and fused silica, $C_3 = 4.94\cdot10^{-49}\, \mathrm{J}\cdot \mathrm{m}^3$ for the $5S_{1/2}$ state, and $7.05\cdot10^{-49}\, \mathrm{J}\cdot \mathrm{m}^3$ for $5P_{3/2}$ \cite{Grover2015,Safronova2011}. The Casimir-Polder correction is negligible for atoms near the surface. We approximate the nanofiber as an infinite dielectric plane when calculating the van der Waals potential~\cite{Alton2011,Stern2011,Frawley2012}. The infinite-plane approximation is accurate to within 20\% for atom-fiber distances less than 200 nm~\cite{LeKien2004,Minogin2010}. For a more detailed discussion see Ref. \cite{Grover2015a}.

%-------------------------------------------------------------------------------------------------------------------
\subsection{Atomic density near the ONF surface}
\label{subsec:density}

The attractive potential given by Eq.~(\ref{eq:vdWCP}) accelerates MOT atoms as they approach the ONF.
As a result of the increased speed of atoms near the surface, the atomic density of untrapped atoms decreases near the ONF. We follow Ref. \cite{Grover2015a} to find the spatial distribution of the atoms. We assume that the system is in equilibrium and that its satisfies energy conservation and the ideal gas law. Under these considerations the attractive potential produces a spatial density distribution
\begin{equation}
\label{eq:density}
\rho(r) \propto \frac{1}{1-U(r)/E}\,,
\end{equation}
where $E = k_{\mathrm{B}} T /2 $ is the total energy of the atom defined far from the ONF surface, $k_{\mathrm{B}}$ is the Boltzmann constant, $T$ the absolute temperature, and $U(r)$ is the surface potential from Eq. (\ref{eq:vdWCP}). This problem is carefully studied in Ref.~\cite{LeKien2008b}, based on quantum-mechanical scattering of atoms off of the surface potential, and Eq. (\ref{eq:density}) produces a similar answer when taking the classical limit of those results.

%-------------------------------------------------------------------------------------------------------------------

\begin{figure}
\centering
\includegraphics[width=1.0\columnwidth]{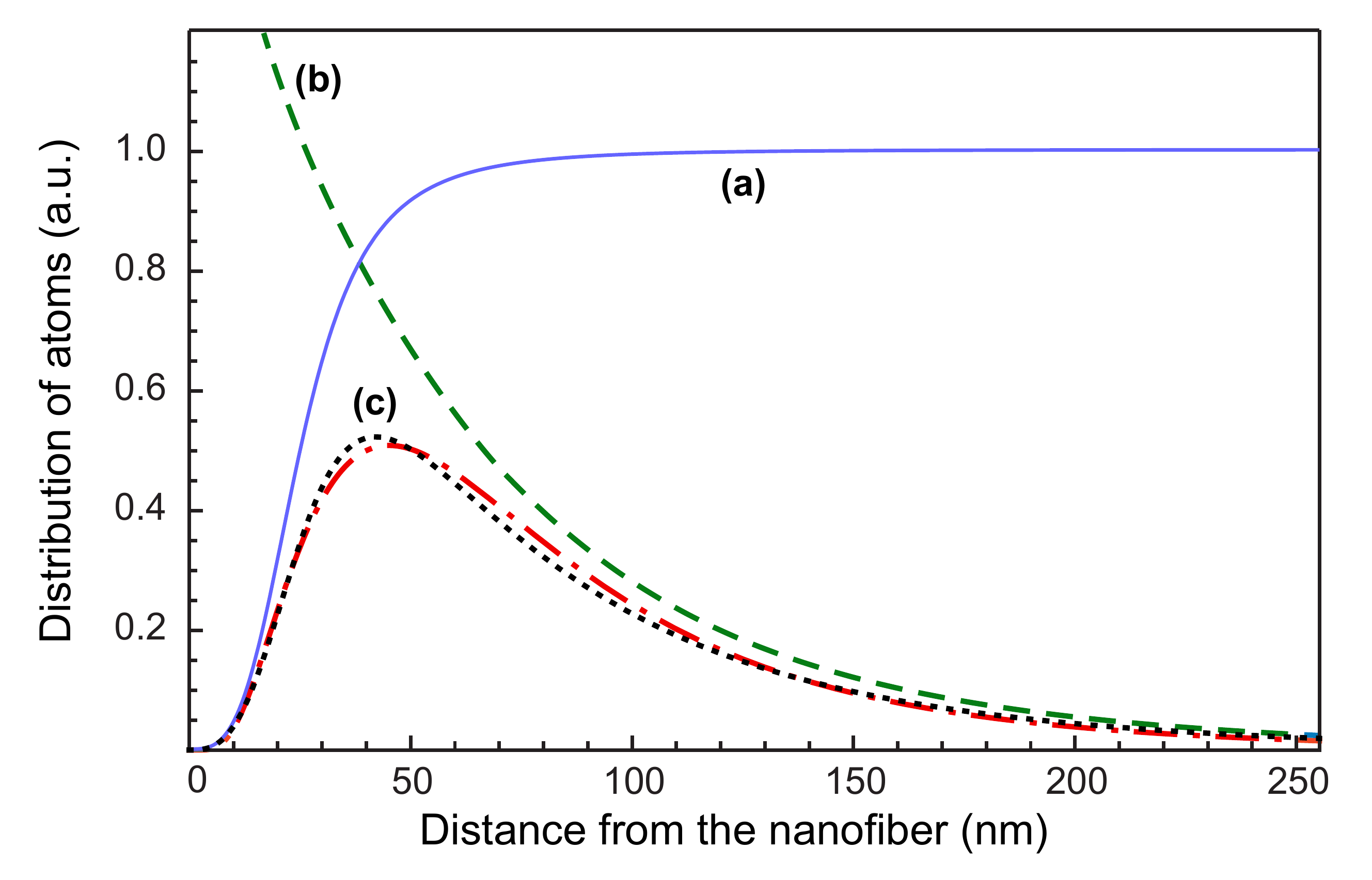}
\caption{\label{fig:posdensity} Distribution of atoms coupled to the ONF. (a, continuous, blue) Predicted normalized position density $\rho$ of cold rubidium atoms ($T = 200\,\mu$K) as a function of atom-surface distance from a fused silica surface. (b, dashed, green) Emission enhancement parameter (in arbitrary units) proportional to the probability of coupling  spontaneously emitted light from a given atom into the ONF. (c, dotted, black) Probability distribution of detecting spontaneously emitted light from a cloud of cold atoms surrounding the nanofiber, obtained from the combination of (a) and (b). (c, dot-dashed, red) Log-normal distribution, adjusted to match dotted line for comparison.}
\end{figure}

Figure~\ref{fig:posdensity} shows the distribution of atoms interacting with the ONF guided mode when considering: (a) the available MOT density from Eq. (\ref{eq:density}), and (b) the radial dependence of the enhancement parameter from Eq.~(\ref{eq:coupling}) (with the same radial dependence than Eq.~(\ref{eq:intensity-ap})). The dotted black line (c) in Fig.~\ref{fig:posdensity} shows the distribution of atoms that contribute to the transmission spectrum. It is the result of the normalized product of (a) and (b), interpreted as the probability distribution of coupling an atom from the MOT to the guided mode of the ONF. The dot-dashed red line in (c) compares the obtained atom distribution to that of a log-normal process, which is a continuous probability distribution produced by the product of independent strictly positive random variables. We see there is a sharp decrease in density near the surface, as expected. This result is in agreement with the numerical calculations of Sagu{\'e} {\it et al.} in Ref.~\cite{Sague2007}.

%================================================================
\subsection{van der Waals bound states}
\label{subsec:vdWstates}

\begin{figure}
\centering
\includegraphics[width=1.0 \columnwidth]{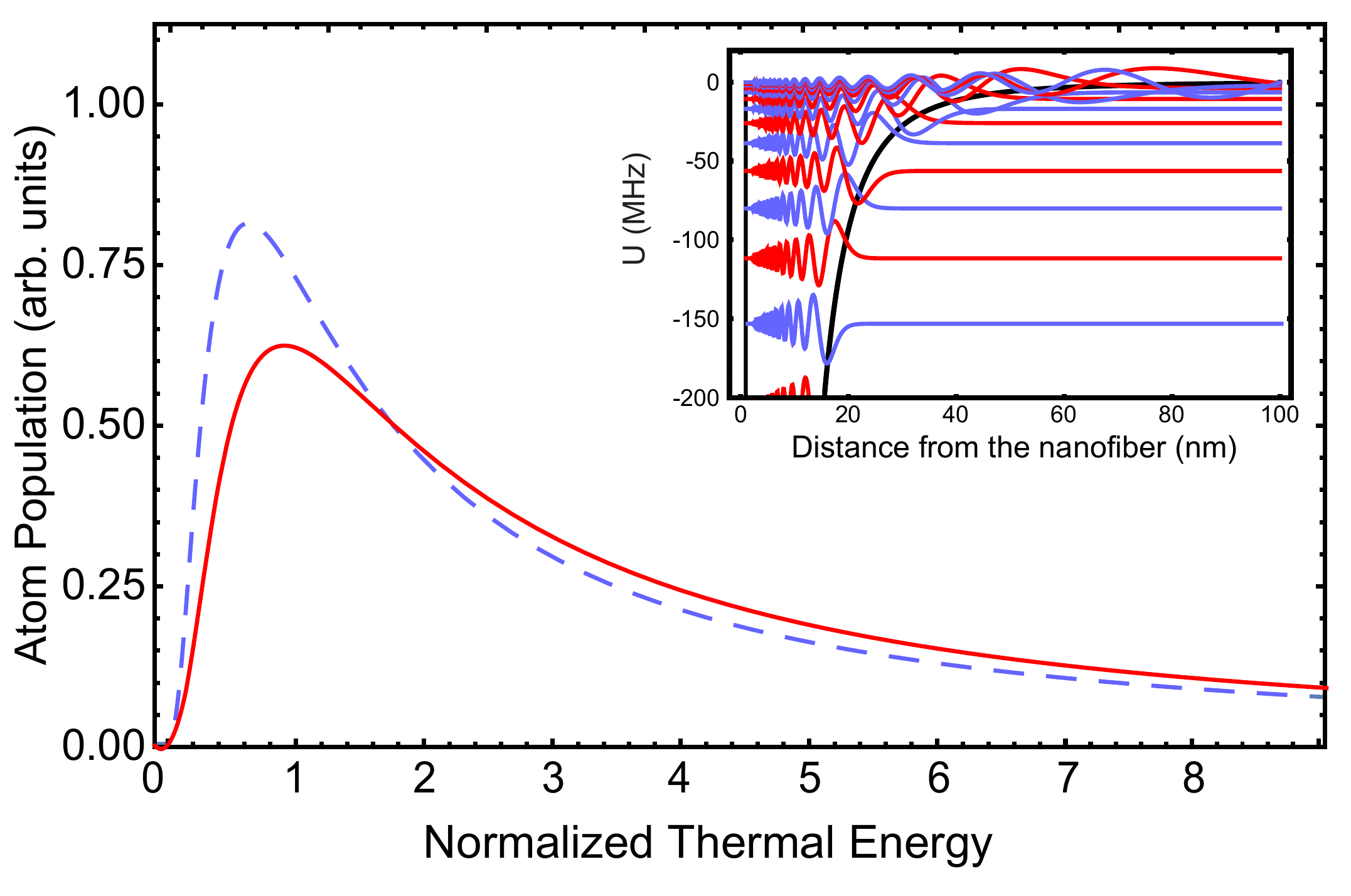}
\caption[vdW Bound State Wavefunctions]{Relative population of atoms in van der Waals bound states between 0 to -20$\Gamma_0$ as a function of the thermal excitation energy normalized to an energy of 20$\Gamma_0$. The population of the bound states follows a Maxwell-Boltzman thermal distribution. The solid red and dashed blue lines are obtained from the wavefunction calculation and from a classical number density model respectively. The insert shows the radial wavefunctions of the atomic bound states in the van der Walls potential.}\label{fig:BoundState}
\end{figure}

Atoms that fall on the ONF can get trapped in the attractive van der Waals potential after an inelastic collision, occupying a bound state of the system. When atoms are at the bottom of the trapping potential, their energy levels are so largely shifted, that for practical purpose we would not see them using near resonant light. However, when higher energy bound states are populated the atomic wavefunction extends far enough from the ONF that the atoms could easily absorb near resonant light, modifying the probed local density distribution. The total density distribution of atoms around the ONF, $\rho_{\text{tot}}(r)$, depends upon the radial distribution of the wavefunction of the populated bound states. The bound states can be calculated numerically \cite{Colbert1992,LeKien2007}. The inset of Fig. \ref{fig:BoundState} shows an example of the calculated wavefunction of high energy bound states for atoms in the electronic ground level. The atomic density distribution near the nanofiber is modified by atoms in van der Waals bound states. This modification will have the spatial dependence of the norm of the bound states wavefunctions and its amplitude is set by the number of occupied bound states. The spatial dependence of the square of the norm of the wavefunctions of bound states scales as $r^{-3/2}$ for a potential of the form $r^{-3}$. This allows us to approximate the total density distribution of atoms around the ONF by adding a term to Eq. (\ref{eq:density}) that captures the contribution of the density of atoms in excited bound states of the surface potential:
\begin{equation}
\label{newdensity}
\rho_{\text{tot}}(r) =\rho(r)+u_{0}r^{-3/2}.
\end{equation}
where $u_{0}$ is a scaling parameter with dimensions of square root of density and it is promotional to the number of atoms in the bound states that contribute to the transmission spectrum signal.

Bound states with trapping frequency much larger than the atomic decay rate would not contribute to the measured spectrum, since the probability to undergo an inelastic collision with the ONF surface during an atomic lifetime is large.  We can estimate a lower limit for the bound states that will contribute to the transmission spectrum signal. If the efficiency of inelastic collision is 5\%, the energy of the lowest bound state that would on average contribute to the signal is above $20\Gamma_0$ ($-120$ MHz). Bound states (displayed in the inset of Fig. \ref{fig:BoundState}) at the very top of the potential will contribute to the measured signal. Although this is a rough estimate, considering a different lowest energy bound state that contributes to the signal does not change the physical behavior nor the qualitative results that we describe. When the temperature of the ONF raises, higher energy bound states get thermally populated. For a large temperature atoms might eventually escape the van der Waals potential. Due to the short interaction time, unbounded states do not contribute to the spectrum. These two issues, the inelastic collisions rate and the excitation of continuum states, impose a lower and an upper limit to the set of bound states that contribute to the measured signal.

We assume a Maxwell-Boltzman distribution of atoms among the bound states, determined by the temperature of the ONF. From the calculation of the atomic wavefunction of the bound states we find that the ONF temperature only determines the amplitude -not the spatial dependence- of the density distribution, which allows us to use $u_{0}$ as a scaling parameter (see Eq. (\ref{newdensity})). Fig.~\ref{fig:BoundState} with a solid red line shows the resulting atomic population as a function of the thermal excitation energy. As the temperature rises, so does the density of atoms populating higher energy bound states. After a threshold temperature, atoms are excited to continuum states, leaving the trap and decreasing the density of atoms. This qualitative semi-classical result is compared with a classical model for number density of desorbed alkali atoms \cite{Lide2003}, shown with a dashed blue line in Fig.~\ref{fig:BoundState}. The functional form of the number density as a function of temperature is
\begin{equation}
u_0(T)\propto\frac{T_{max}}{T}\exp{\left(1-\frac{T_{max}}{T}\right)}
\label{u(T)}
\end{equation}
where $T_{max}$ is a threshold temperature at which the number density reaches its maximum value, and for rubidium in glass is the high temperature of $T_{max}=9705$ K  \cite{Lide2003}.

The spectroscopy of bound states can be very challenging, since bound atomic ground states can couple to several bound atomic excited states. The transition strengths between different bound levels are given by the Franck-Condon factors, and they have been calculated for atoms in the van der Waals potential of an ONF in Ref. \cite{LeKien2007b}. However, as we discussed above, we are interested in measuring a set of bound states at the very top of the van der Waals potential. In this region, most of the levels are too close together to be spectroscopically resolved. Moreover, atoms in high energy bound states move slow enough (low trapping frequency) that we can use the quasistatic theory of line broadening to describe their spectroscopic signal \cite{Hedges1972,Julienne1992}. In this limit, the spectrum is given only by the local potential felt by the atoms. Calculating the atomic energy shift as a function of position (obtained from Eq. (\ref{eq:vdWCP})) and the atomic density distribution (see Eq. (\ref{newdensity})) is enough to determine the spectroscopic signal of the sample \cite{Russell2009}. 

Here we only consider the case where the atomic motion is restricted to the radial direction. Non-radial atomic trajectories can be deflected by the van der Waals potential, to the point that atoms moving fast enough can orbit the ONF. This angular momentum due to deflected atomic trajectories can produce a centrifugal potential barrier that will modify the bound state of the system. However we are not able to resolve these small corrections in our experiment. Atoms from the MOT are too cold to undergo a close orbit around nanofiber, almost directly falling onto the nanofiber, as found by numerical simulations \cite{Grover2015,Sague2007}, and atoms in the van der Waals potential are assumed to exchange momentum with the ONF only perpendicular to the interaction surface.

%================================================================

\subsection{Probability of light absorption}
\label{subsec:shifts}

The atom-surface potential produces a state-dependent shift of the atomic levels that varies as a function of the position of the atoms.
The shifts produce a spatially-varying absorption spectra, where the probability of absorption by an atom at position $r$ is \cite{Foot2005}:
\begin{equation}
p_{\mathrm{abs}}\left( r \right) \propto \frac{1}{1+s+4\left( \frac{\delta_{\text{vdW}} \left( r \right) +\delta_{\text{L}}}{\Gamma} \right)^2}\,,
\label{eq:shift}
\end{equation}
where $s = I/I_{\mathrm{sat}}$ is the saturation parameter ($I_{\mathrm{sat}} = 3.58\, \mathrm{mW}\cdot\mathrm{cm}^{-2}$ for a uniform sublevel population distribution~\cite{Steck2001}), $\delta_{\text{L}} =\left( \omega_{\mathrm{L}} - \omega_0\right)$ is the detuning of the driving (i.e. probe) from free space atomic resonance and $\delta_{\text{vdW}} \left( r \right) = \left( U_e(r) - U_g (r)\right )/\hbar$ is the atom-surface shift assuming a two-level atom with contributions from the excited $U_e$ and ground $U_g$ levels (see Eq. (\ref{eq:vdWCP})). Since $C_3$ is larger for the excited state than for the ground state, the van der Waals potential produces a red shift on the atoms near the surface of the ONF.

%================================================================

\subsection{Model of the resonance spectrum}
\label{subsec:model}

The transmitted intensity, for low number of atoms, is proportional to one minus the probability of absorption of a photon from the nanofiber mode. The absorption spectrum is found by the spatial average of the probability of a guided photon being absorbed by an atom in the cloud (Eq.~(\ref{eq:shift})), the atomic density distribution (Eq.~(\ref{newdensity})) and the intensity of the field at the position of the atoms (Eq.~(\ref{eq:coupling})), shown unnormalized as:
\begin{equation}
P_{abs}(\omega) = \int r \mathrm{d}r p_{\mathrm{abs}}(r,\omega) \rho_{\text{tot}}(r)\alpha(r),
\label{spectrummodel}
\end{equation}
which represents the convolved spectrum for the steady-state density distribution.

%=================================================================

\section{Experiment}
\label{sec:expt}

\begin{figure}
\centering
\includegraphics[width=0.9 \columnwidth]{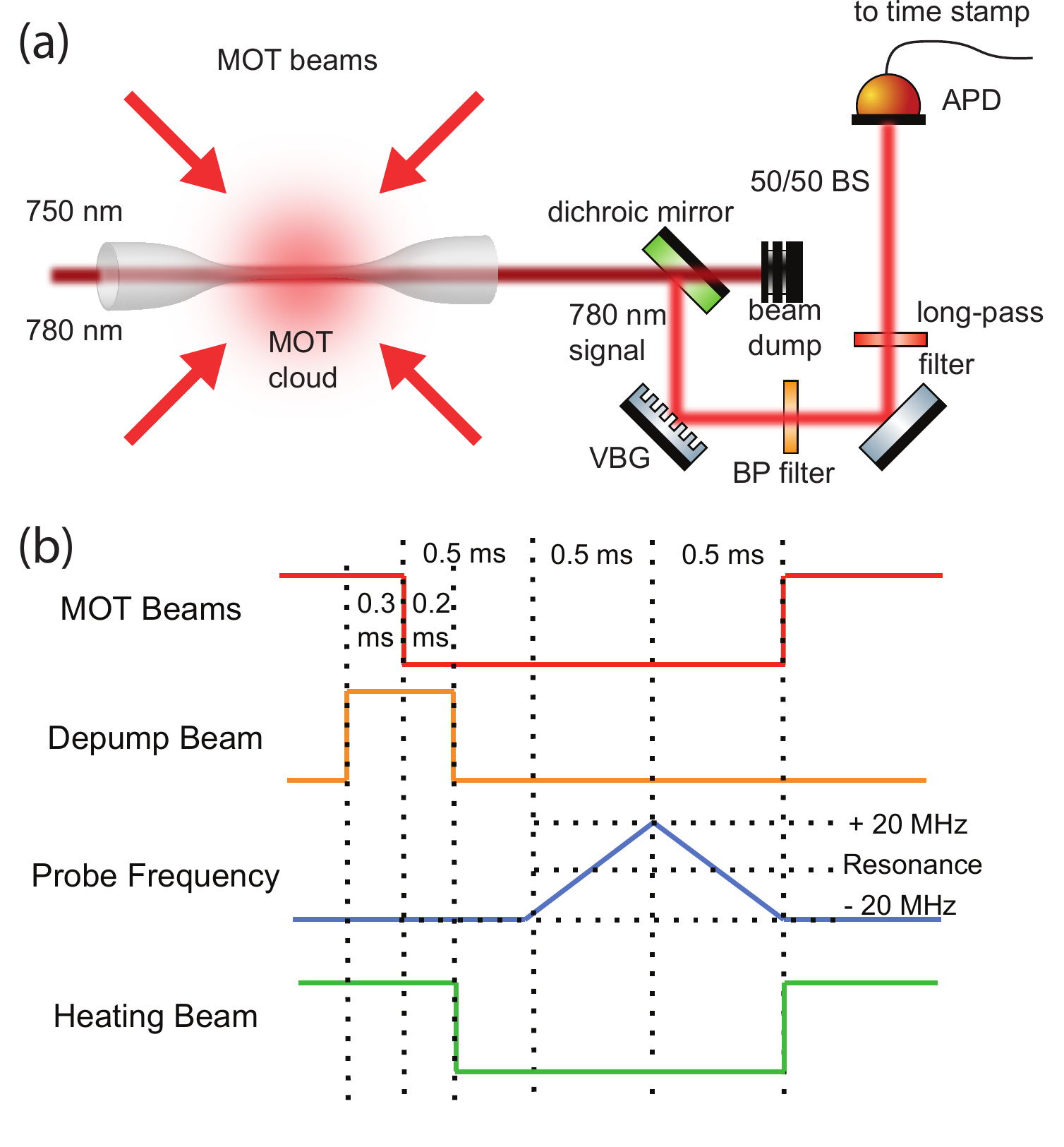}
\caption[Experimental schematic for spectrum measurements]{\label{fig:schematic} Experimental schematic and time cycle (a) A 780 nm resonant probe beam launched through the nanofiber scans around the atomic resonant frequency and a APD measures the transmitted light. A 750 nm laser also launched through the nanofiber heats it. The MOT forms around the nanofiber The transmitted signal is filtered by a volume Bragg grating (VBG), bandpass (BP) filter, and long-pass filter before being sent to an APD whose output is time stamped for processing. (b) Experimental cycle that includes the trapping (MOT Beams), preparation  of the sample, probe (780 nm) and heating beam (750 nm).}
\label{fig:setup}
\end{figure}

The ONF overlaps with a $^{87}$Rb MOT (see Fig. \ref{fig:setup} (a)). The MOT, with a typical magnetic field gradient of 10 G/cm, is loaded from a background vapor, generated with a Rb dispenser, and produces a cloud of approximately $10^8$ atoms at temperatures of few hundreds of $\mu$K \cite{Grover2015}. We know that the fundamental mode transmission is greater than 95\%, based on our fiber-pulling reproducibility \cite{Hoffman2014a}.
We estimate the diameter of our ONF to be $240\pm 20$ nm, with a uniformity better than $\pm$ 10 nm over a length of 7 mm.

We measure atomic absorption with a weak, near-resonance
beam (780 nm) coupled through the ONF, counting the transmitted photons with avalanche photodiodes (APD).
Because optical powers near 10 pW saturate the APDs, great care
must be taken to filter stray light and maintain low probe
power, detailed in Ref. \cite{Grover2015}. TTL pulses from the APDs are counted with a PC time-stamp card with 164 ps resolution. The light intensity of the probe is kept to less than one tenth of the saturation intensity for the transition to make sure there is no power broadening. Background counts from all sources (such as ambient light or incomplete MOT turnoff) are negligibly low. Our usual counting rates are $7 \times 10^4$~s$^{-1}$, well below saturation of the APDs and corresponds to a transmitted probe power of $19\times10^{-15}$ W. 

We use saturation spectroscopy to lock the probe laser to an outside reference. This gives us a knowledge of the frequency location of the $5S_{1/2},F=2\rightarrow 5P_{3/2},F'=3$ transition within less than one MHz. The frequency of the probe is scanned across resonance (see Fig. \ref{fig:setup} (b)) with a double pass AOM to obtain the transmission spectrum of the atoms. 
%-------------------------------------------------------------------------------------------------------------------
The transmission spectrum of MOT atoms coupled to the ONF guided mode is symmetric. Due to the low density of atoms near the nanofiber surface (see Fig.~\ref{fig:posdensity}), few atoms are close enough to the surface to experience a significant shift of their energy levels \cite{Sague2007}. However, by shining a far off resonance (750 nm) laser collinear with the probe the ONF heats (see Fig.~\ref{fig:setup}), thermally exciting the atoms physisorbed on the surface. This creates a larger distribution of atoms around to the ONF surface that can be probed.
%-------------------------------------------------------------------------------------------------------------------

Figure~\ref{fig:setup} (b) shows the basic timing of the experimental cycle. We control the number of atoms in thermally excited bound states by changing the power of the 750 nm heating beam. The ONF maintains temperature during the short probing period. We control the number of atoms from the MOT using a free space depumping beam that moves atoms to the $F=1$ hyperfine manifold of the ground state. With this method the optical transmission is independent of MOT density, since most of the atoms in the MOT are in the $F=1$ state that is transparent to the probe.  The redundancy of the probe scan, which goes from red to blue detuned frequency and then from blue to red detuned frequency (see Fig. \ref{fig:setup} (b)), provides a test to rule out transient effects such as increase or decrease of the average number of atoms or changes in atomic density distribution during the interrogation time.

To compensate for systematic fluctuations and slow drifts of the probe power, each spectrum is processed from two independent datasets. A ``signal'' set collected with the MOT active, and a ``background'' set collected with the MOT turned off. We calculate the absorption spectrum by taking the appropriate ratio of the two datasets.

%-------------------------------------------------------------------------------------------------------------------
\section{Results}
\label{sec:res}
%-------------------------------------------------------------------------------------------------------------------

\subsection{Measured spectra}
\label{subsec:spectra}

\begin{figure} 
\centering
\includegraphics[width=1\columnwidth]{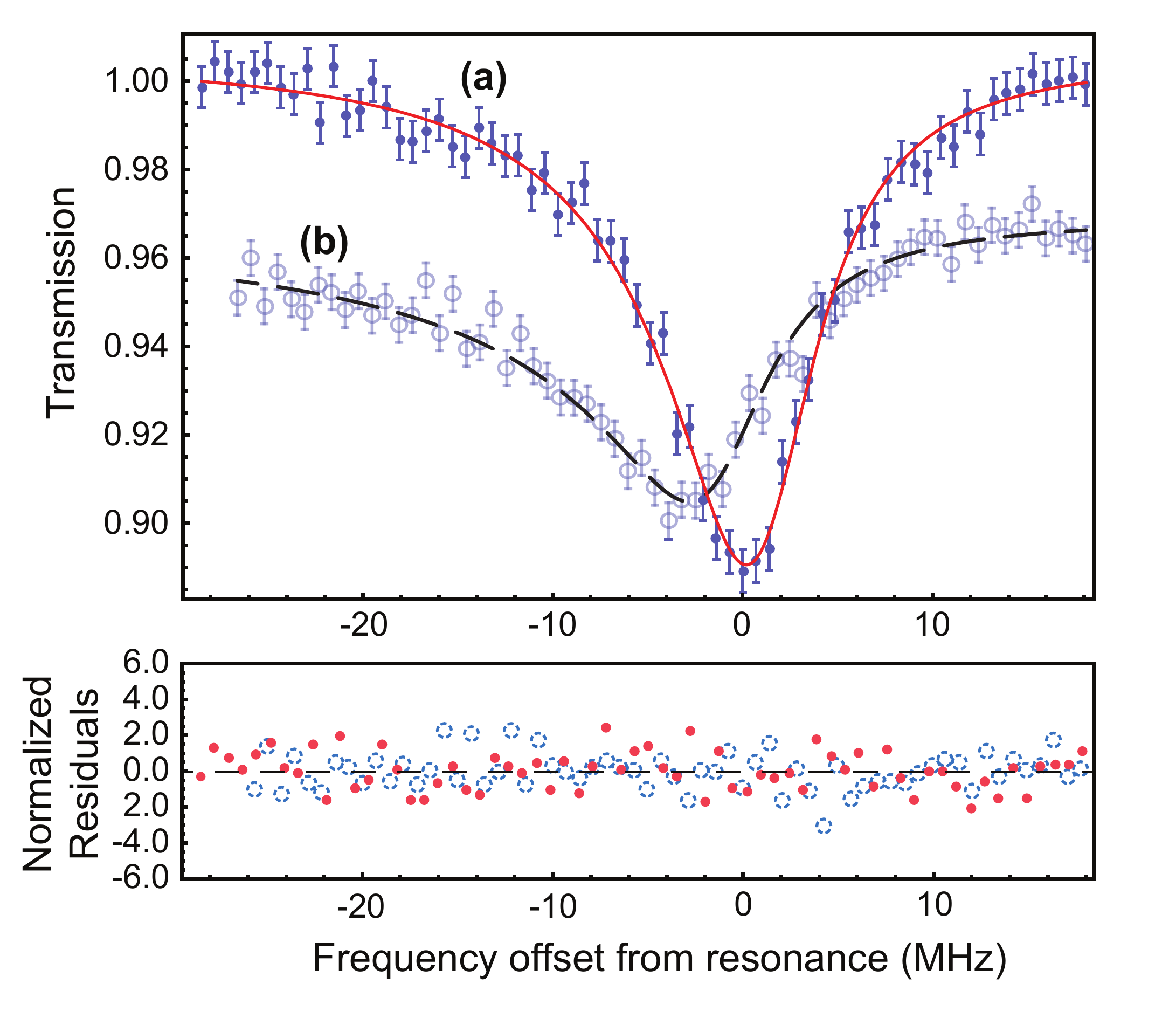}
\caption[Symmetric transmission spectrum]{\label{fig:spectra}Transmission spectrum and fits to Eq.~(\ref{eq:spectrumfit}) through the nanofiber with no heating (a, solid red line, filled circles) (0~$\mu$W of 750~nm power) and with heating (b, dashed black line, open circles) (250 $\mu$W of 750~nm power for 15~ms). The error bars are statistical. The lines are the fit to the model, with the normalized residuals shown below.The reduced $\chi^{2}$ is 1.11 (a) and 1.32 (b).}
\end{figure}

Figure \ref{fig:spectra} shows examples of the transmission spectra through the ONF showing optical absorption on the $5S_{1/2}, F=2 \rightarrow  5P_{3/2}, F'=3$ transition of $^{87}$Rb. The accumulation time for a spectrum is about 400 seconds. This corresponds to roughly 19,000 data-collection periods.

Modest variations of the dispenser current, under our operating conditions, do not influence the number of accessible atoms near the ONF, in contrast to the observations in Ref.~\cite{Nayak2012a}. This suggests that we are in a regime where the density of atoms near the ONF is only limited by the heating power of the 750 nm desorption laser and not the accumulation rate of atoms on the ONF. 

Figure  \ref{fig:spectra}(a) comes from MOT atoms only. The distribution has very little asymmetry and a full width at half linewidth of 8.9$\pm$0.2 MHz, with an error determined by the standard error of the fit. In Fig.~\ref{fig:spectra}(b), the ONF is heated by 250 $\mu$W of a 750 nm laser, and shows a long tail on the red side of resonance. In this case, we also observe attenuation across the scan region that we believe to come from atoms at the bottom of the van der Waals potential and/or transit broadening from desorbed atoms rapidly flying away from the ONF. The asymmetry on the red side of Fig.~\ref{fig:spectra}(b) and the same shape on the blue side of Figs.~\ref{fig:spectra}(b) and \ref{fig:spectra}(a) suggests that a percentage of the atoms freed during the desorption process are loaded into excited bound states of the van der Waals potential. We model this by numerically calculating the wavefunctions of atoms in thermally populated van der Waals bound states, as seen in Fig.~\ref{fig:BoundState}.

Adding to the cycle a 5~ms period of letting the atoms fall on the ONF surface has a negligible impact on the degree of asymmetry of the spectrum. This indicates that the accumulation and desorption processes reach steady state in this regime. 

\subsection{Fitting Model}

The fitting function for spectra as those in Fig.~\ref{fig:spectra} is obtained from Eq. (\ref{spectrummodel}), but it includes an overall scaling ($I_{0}$), an amplitude offset $(O)$, and frequency translation $(\omega_0)$ parameters which, although constrained, allow the fit to accommodate changes in the laser frequency and intensity:

\begin{equation}
\begin{multlined}
\label{eq:spectrumfit}
P_{abs}(\omega)[\omega_{0},\Gamma_{0},u_{0},T_{\text{MOT}},I_{0},O] = \\
O- I_{0}\int [\alpha(r) (\rho(r,T_{\text{MOT}})+u_{0}r^{-3/2})\\
 \times p_{\mathrm{abs}}(r,\omega-\omega_{0},\Gamma_{0})]r\mathrm{d}r\,
\end{multlined}
\end{equation}

\vspace{1em}

We perform a sensitivity analysis on every parameter in the model. The quantification of the effect of $u_{0}$ on the spectrum is the primary motivation for the model, and we can establish a relationship between $u_{0}$ and the power of the heating beam from the bound state calculation (see Eq. (\ref{newdensity}) and Fig. \ref{fig:BoundState}). The absolute temperature scale is difficult to establish without precise knowledge of the very-near-surface potential, namely the interatomic spacing at the heterogenous silica-Rubidium interface. The role of $\Gamma_{0}$ and its deviation from the natural linewidth, however, is more subtle \cite{Barnett1996}. $\Gamma_{0}$ enters in $p_{abs}$, the resonant absorption component of the model. Variations in this parameter can have significant effects, and we consistently measure a 2~MHz increase from the natural linewidth which we do not yet understand. Doppler broadening of this size corresponds to a temperature of 72~mK, which is too high compared to the measured value \cite{Grover2015a}. Collective effects of atoms along the ONF, such as superadiance, can also increase the radiative decay rate \cite{Solano2017a}, contributing to a broader linewidth. However, this effect has a linear dependence on the atom number and it can be easily tested. Variations in the MOT density do not strongly affect the measured width, meaning that we have densities low enough that collective effects are negligible. A possible explanation could be the modification of the atomic linewidth due to Purcell effect of the ONF, but measurements for similar atomic distribution have shown this to produce only a 10\% increase \cite{Solano2017b}. Sagu{\'e} {\it et al.} \cite{Sague2007} measure a similar broadening in Cs, on the order of 20\%. We do expect a significant background of hot continuum-state atoms, but their Doppler-broadened linewidth would be on the order of hundreds of MHz and should manifest as broadband background. The presence of a magnetic field gradient introduces the possibility of Zeeman broadening, but the linewidth does not respond to changes due to magnetic field from the MOT coils. 

The temperature of the atoms in the MOT, $T_{\text{MOT}}$, affects the steady-state component of the the atomic distribution, $\rho$ (see Eq. (\ref{eq:density}) and Fig. \ref{fig:posdensity} (a)). Decreasing the temperature extends the low-density region, and increasing it brings the distribution closer to the ONF. $T_{\text{MOT}}$ is negligible for high-asymmetry spectra because the distribution is dominated by the desorption component, but is very sensitive at minimum asymmetry.

It is important to note that the influence of  $\Gamma_{0}$ and $T_{\text{MOT}}$ on the fit is distinct and between them they do not show covariance. They converge to their true values when both are left as free parameters, and when either is fixed.

\subsection{Asymmetry parameter}
\label{subsec:asy}

We calculate the spectral asymmetry by integrating the absorption signal to the blue (red) side of resonance, giving us a quantity defined as $R$ ($L$). The asymmetry parameter $A$ is then:
\begin{equation}
A=\frac{L-R}{L+R}
\label{eq:assym}
\end{equation}

This function is zero for a symmetric transmission spectrum and can increase (decrease) to a maximum (minimum) of +(-) 1. Because van der Waals produces red-biased shifts, we expect this number to be strictly positive. After fitting the numerical model, we evaluate the same parameter by integrating over the regions rather than calculating a discrete sum.

\subsection{Analysis of the fits}

Table \ref{tab:paramtable} shows the results of the fitting for the physically significant parameters of the model (see Eq.~(\ref{eq:spectrumfit})). The asymmetry parameter (see Eq.~(\ref{eq:assym})) depends on the relative proportion of atoms in bound state to MOT atoms, given by $u_{0}$. The asymmetric lineshape becomes visible in the $u_{0} < 1$ range. At higher values ($u_{0}>5$), the density distribution $\rho_{\text{tot}}$ quickly becomes dominated by the $u_{0}$ term (see Eq. (\ref{newdensity})). Since $\rho_{\text{tot}}$ is not normalized, in the limit when $u_{0}$ dominates and $\rho(r)$ is neglected the fits of the transmission spectrum become almost insensitive to the actual value of $u_{0}$. Values of $u_{0}$ can become very large ($u_{0}>1000$). In this limit, the spectral asymmetry begins to saturate. At higher power of the heating beam, atoms can escape the van der Waals potential and $u_{0}$ becomes smaller again. The reduced $\chi^{2}$ of the fits ($\chi_{r}^{2}$) seems to increase with the asymmetry parameter. There may be effects of the desorption mechanism that we have not accounted for in the model.

\begin{table}[H]
\begin{center}
    \begin{tabular}{| c | c | c | c | c | c | c |}
    \hline
    Power   & $\omega_0/2\pi$ & $\Gamma_0/2\pi$ &$T_{MOT}$  &{$u_{0}$} & $A_{calc}$ &$\chi_{r}^{2}$ \\
    ($\mu$W) &(MHz)&(MHz)&($\mu$K)&&&\\ \hline
    0 & 5.9$\pm$0.2 & 8.1$\pm$0.3 & 332$\pm$17 & {0 (f)} & 0.14 & 1.11 \\ \hline
    40  & 0.7$\pm$0.1 & 8.1 (f) & 336$\pm$23 & {0.19$\pm$0.09} & 0.19& 1.16\\ \hline %0.02   0.19 $\pm$ 0.09
    120 & 1.0$\pm$0.2 & 9.2$\pm$1.0 & 332 (f) & {7182$\pm$269}& 0.36 & 1.91\\ \hline %19.97  7182 $\pm$ 269
    250 & 0.9$\pm$0.3 & 8.4$\pm$0.9 &  332 (f) &  {5897$\pm$612}& 0.26 & 1.32\\ \hline %3.8 5897 $\pm$ 612
    350 & 0.8$\pm$0.2 & 9.5$\pm$2.4 & 332 (f) &  {0.11$\pm$0.11} & 0.12 & 1.29\\ \hline %0.01 0.11 $\pm$ 0.11
    \end{tabular}
    \caption{Best-fit parameters for each of the measured spectra. Fixed parameters are indicated with an (f).}
    \label{tab:paramtable}
\end{center}
\end{table}

%-------------------------------------------------------------------------------------------------------------------

The experimental errors are dominated by the uncertainty in determining the center of the resonance. We vary the position by plus and minus one bin (about 500 kHz) and recalculate the value of $A$ to set the limits of the error. The uncertainty due to the counting statistics is negligible on this scale. 

\begin{figure}
\centering
\includegraphics[width=1.0 \columnwidth]{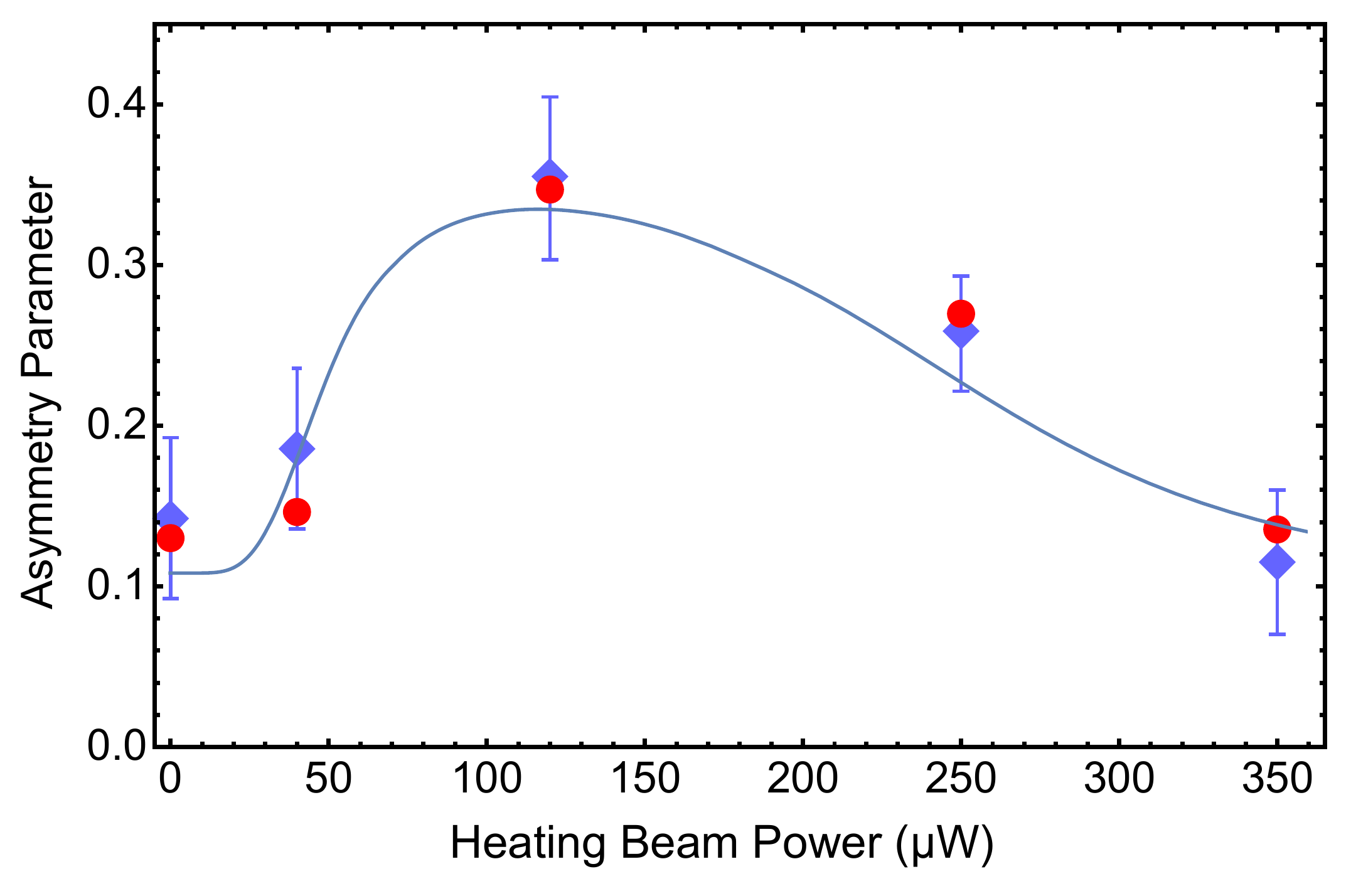}
\caption[Asymmetry temperature dependence]{Lineshape asymmetry $A$ as a function of heating laser power. Blue points (diamonds) are extracted from data, and red points (circles) are calculated from fits to the spectrum model. See Table \ref{tab:paramtable} for parameters. The continuous blue line comes from the asymmetry parameter of simulated spectra calculated from the numerically modeled of the thermally populated bound states as explained in Sec. IV E.}\label{fig:asymmetry}
\end{figure}

By measuring the asymmetry parameter induced by the van der Waals potential, we can see that the density of atoms in the excited bound states depends on the temperature of the ONF that gets transferred to the atoms, shown in Fig.~\ref{fig:asymmetry}. If the atoms are too cold (no 750 nm power) they do not have sufficient kinetic energy to overcome the van der Waals potential. If the ONF is too hot (high 750 nm power) the atoms fly away before they can interact with the near resonant probe. There is a power that maximizes the number of atoms in the excited bound states of the van der Waals potential, maximizing the asymmetry. This optimum ONF temperature (or heating beam power) balances atoms in high energy bound states that interact with the probe for sufficiently long time before escaping from the potential or undergoing inelastic collisions with the ONF surface. In our configuration, we found that this power is approximately 120~$\mu$W. It is difficult to directly relate this power to an actual temperature of the ONF, because we do not know all the dissipation mechanisms or the proportion of optical power lost to heat in the ONF.

\subsection{Comparison with a phenomenological model}

We developed a phenomenological model of spectral asymmetry versus heating power to understand our experimental data. Asymmetry is calculated according to Eqs. (\ref{eq:spectrumfit}) and (\ref{eq:assym}) as a function of desorption density $u_{0}$ from Eq. (\ref{newdensity}). The core of this model is in establishing the relationship between heating power and $u_{0}$. The behavior of $u_{0}$ for different fiber temperatures $T$ is discussed in Sec. \ref{subsec:vdWstates} and shown in Fig. \ref{fig:BoundState}. We incorporate the relation $T \propto P^{{1}/{g}}$ as an ansatz, with $g$ being a fitting parameter and $P$ the optical power of the heating beam. The power law that relates temperature with power is well know from the problem of black body radiation, where $g=4$. Because the ONF is smaller in size than the wavelength radiated, the power law deviates from Plack's law. In our case $g$ is expected to be less than 3 in accordance with the results of \cite{Wuttke2013a}. 

An unexpected characteristic of our system is that $u_{0}$ has an additional dependence on the power of the heating beam. Because the heating beam is blue detuned, its evanescent field produces a repulsive potential which acts on nearby atoms. For an atomic cloud of temperature $T_{MOT}$, the number of atoms $N$ that can reach the ONF surface is given by
\begin{equation}
\begin{split}
N=\int_{V_{b}(P)}^{\infty} \sqrt[]{\frac{m}{2\pi k_{B}T_{MOT}}}\exp{\left(-\frac{m V^{2}}{2k_{B}T_{MOT}}\right)}dv\\
\propto(1-\text{Erf}[b_o\sqrt[]{P}])\\
\label{N}
\end{split}
\end{equation}
%\begin{equation}
%\begin{split}
%N = \int_{V_{b}(p)}^{\infty} \sqrt[]{\frac{m}{2\pi k_{b}T}}e^{-\frac{m V^{2}}{2k_{b}T}}dv\\
%=(1-\text{Erf}[\frac{V_{b}(p)}{k_{b}T}])\\
%=A_o(1-\text{Erf}[\beta\sqrt[]{p}])
%\end{split}
%\end{equation}
where $P$ is the heating beam optical power and $b_o$ depends on the repulsive dipole potential $V_{b}(P)$ produced by the blue-detuned heating beam, as $b_o=\sqrt{V_{b}(P)/k_b T_{MOT}}$. This can be calculated knowing the atomic polarizability and the effective guided mode area \cite{Sague2007}. Considering the D1 and D2 line polarizability of $^{87}Rb$, and a MOT temperature of 330 $\mu$K, we obtain a value of $b_o=0.142$. This phenomenon limits the number of atoms on the ONF surface that can be thermally excited, reducing the asymmetry of the spectrum for large heating beam power.

We can calculate the asymmetry parameter for simulated data, considering Eq. (\ref{spectrummodel}) and an atomic density distribution given by  Eq. (\ref{newdensity}), where $u_0$ depends on the heating beam power. This dependence will be given by the Maxwell-Boltzman distribution shown in Fig. \ref{fig:BoundState} (see Eq. (\ref{u(T)})), which is a function of the ONF temperature as $T \propto P^{{1}/{g}}$, times the number of atoms at the nanofiber surface that can be thermally excited, given by Eq. (\ref{N}). If we want to compare this model to our data there are two free parameters to manipulate: an over all amplitude and the exponent $g$. The blue solid line in Fig. \ref{fig:asymmetry} is a fit to the data points. The obtained exponent is  $g=2.26\pm 0.05$, in agreement with \cite{Wuttke2013a}. Allowing $b_o$ to vary as a fitting parameter we find $b_o = 0.156 \pm 0.019$ and $g = 2.15 \pm 0.136$, consistent within errors.

Although the fit to the phenomenological model is in good agreement with the data, more statistics or direct temperature measurements are necessary to establish a true quantitative relationship, opening a door to future investigations.

\section{Discussion}
\label{sec:disc}

A simple picture emerges from our phenomenological model that may guide future realizations of optical trapping using surface potentials such as the van der Walls \cite{Chang2014}. The atoms are adsorbed (bound) from the cold atomic cloud, the heated nanofiber can provide energy to place a few atoms in the top levels of the potential from where we can observe them near resonance. If the temperature it too large, the previously adsorbed atoms get desorbed, leaving the surface. In this case we expect to see no asymmetry. Just the right temperature succeeds in populating the higher bound states. This picture also explains the decrease in the transmission that we observe without any sharp spectral features on the asymmetric traces such as Fig.~\ref{fig:spectra} (b). In our case the number of atoms in higher energy bound states of the van der Waals potential was not limited by the temperature of the ONF but by the repulsion created from the laser used to heat the nanofiber. Further experiment might benefit from independent control of initial atom number and the nanofiber temperature.

%=================================================================
\section{Conclusions}
\label{sec:concs}
We presented the transmission spectra of $^{87}$Rb atoms around an optical nanofiber. When using a cloud of cold atoms from a MOT the spectra are nearly symmetric, which we interpret as the atoms being too far from the surface of the nanofiber to produce significant broadening due to the van der Waals potential. By heating the atoms deposited on the surface of the nanofiber it is possible to observe an asymmetry in the spectra, quantify it, and extract information about the density distribution of atoms near the ONF surface. The qualitative agreements between the measurements and a phenomenological model suggests that we can thermally excite atom-nanofiber van der Waals bound states, allowing us to probe them half a millisecond later, while maintaining a constant average transmission. 

Use of this desorption technique could enable future work probing the properties of the atom-nanofiber interaction, and open a new parameter space for applications that operate in the very near surface regime.

\section{Acknowledgements}
We acknowledge discussions with A. Rauschenbeutel on this topic. This work was supported by the National Science Foundation.

\bibliographystyle{osajnl.bst} 
\bibliography{AAMOPreview.bib}
\end{document}